\documentclass[reprint,
superscriptaddress,
 amsmath,amssymb,
 aps,
floatfix,
]{revtex4-1}

\usepackage{graphicx}
\usepackage{dcolumn}
\usepackage{bm}
\usepackage{comment}
\usepackage{hyperref}
\usepackage{xcolor}
\usepackage{soul}
\usepackage{bm}

\begin{document}

\preprint{APS/123-QED}

\title{Spiral defect chaos with intermittency increases mean termination time}


\author{Mahesh Kumar Mulimani}
 \email{mmulimani@ucsd.edu}
 \affiliation{Department of Physics, University of California, San Diego.}
\author{Wouter-Jan Rappel}
 \email{rappel@physics.ucsd.edu}
 \affiliation{Department of Physics, University of California, San Diego.}
\affiliation{Department of Bioengineering, University of California San Diego, La Jolla, San Diego, 92093, CA, United States of America}


\begin{abstract}
Cardiac models are examples of excitable systems and can support stable spiral waves. 
For certain parameter values, however, these spiral waves can become unstable, resulting in 
spiral defect chaos (SDC), characterized by the continuous creation and annihilation of spiral waves
and thought to underlie atrial fibrillation.
During  SDC, the number of spiral waves fluctuates and eventually drops to zero, marking the termination of activity.
In this work, we demonstrate that varying a single parameter allows the system to transition from SDC to a single spiral wave, passing through an intermediate regime of intermittency.
In this intermittent dynamics, intervals of SDC are sandwiched between non-SDC intervals
during which the number of spiral waves remains small and constant. 
We quantify this intermittency and show that the mean termination time 
increases significantly as the control parameter approaches values for which a
single spiral wave is stable. 
In addition, we find that 
it is also possible to have intermittently 
present quasi-stable spiral waves in part of the computational domain while the 
remainder of the domain exhibits SDC.
Our results may have implications for clinical atrial fibrillation, which often shows intermittency, 
switching back-and-forth between fibrillation and normal sinus rhythm.

\begin{description}
\item[Keywords]
\textbf{Cardiac arrhythmias, Chaos, Spiral defect chaos, Intermittency}
\end{description}
\end{abstract}

\maketitle

\section{Introduction}

Spiral defect chaos (SDC) is a ubiquitous dynamical process in spatially extended excitable media that involves the continuous creation and annihilation of spiral waves. It is observed in oscillatory Belousov-Zhabotinksy chemical reactions~\cite{qiao2009defect, beta2006defect}, in fluid dynamics such as the Rayleigh-Bernard convection cells~\cite{morris1996spatio}, and in model systems such as the complex Ginzburg-Landau equation \cite{huepe2004statistics}. In addition to these physical systems, SDC is observed in cardiac tissue where it is believed to underlie dangerous and potentially lethal  arrhythmias \cite{witkowski1998spatiotemporal,kumagai1997simultaneous,chen2000dynamics,davidenko1992stationary,nattel2017demystifying,christoph2018electromechanical,Naretal12b,karma2013physics,rappel2022thephysics}.
For example, atrial fibrillation (Afib), the most prevalent cardiac arrhythmia, is 
thought to be driven by spiral waves that are self-regenerating. During Afib, the electrical activation 
of the heart is disordered. Afib can occur in patients for just a few minutes, after which normal sinus rhythm is restored, or persist for months or even longer at a time. ~\cite{jalife2002mother,rappel2024spatially,wijesurendra2019mechanisms,narayan2012computational}.
This difference in time scales is also reflected in the clinical classification of 
Afib:  
paroxysmal atrial fibrillation (Afib) is defined as an arrhythmia that lasts less than seven days and typically resolves on its own while persistent Afib refers to arrhythmias that last more than seven days and often require medical intervention to restore normal sinus rhythm
 ~\cite{heijman2014cellular,gunawardene2022atrial}. 
 In addition to paroxysmal fibrillation, patients can also 
 exhibit alternating periods between Afib and atrial flutter, characterized by a 
 more ordered activation patterns \cite{tunick1992alternation,clair1993spontaneous,paydak1998atrial}.

The propagation of activation waves in cardiac tissue can be simulated using  electrophysiological models, which
formulate  the membrane voltage of cardiac cells in terms of a set of 
coupled equations \cite{karma2013physics,rappel2022thephysics}. 
These models are excitable and, by breaking the wavefront, it is possible to 
create a single spiral wave.
For certain model parameters, however, this single spiral wave 
is unstable, resulting in the stochastic creation and 
annihilation of spiral waves and SDC \cite{fenton2002multiple}. 
Numerical studies of these models have demonstrated that the  
number of spiral tips in SDC is a stochastic quantity and that its birth-death process 
 can be cast into a master equation,
a commonly used approach in the field of population
dynamics \cite{lande2003stochastic,van1992stochastic,Gard_book09}.
Using this master equation, it is possible to show that there is a finite probability 
of spiral wave termination and that the resulting termination times are 
exponentially distributed. 
This is consistent with the clinical finding that 
fibrillation involves the creation and annihilation of spiral waves with inter-event times obeying a Poisson renewal process   ~\cite{dharmaprani2019renewal}.
Furthermore, numerical studies also showed  that  the mean termination time, $T_{term}$, depends critically on the size of the simulation domain. 
More specifically, it was shown that $T_{term}$ increases exponentially with the domain size.  
This aligns with clinical findings, which show that 
larger hearts have a higher risk for Afib ~\cite{garrey1924auricular,saadeh2024relationship,hartley2021size}.

In this study, we show that increasing the domain size is not the only way 
to increase the mean termination time. We show that, for certain parameter sets, 
a generic cardiac model can exhibit intermittency, which can 
greatly increase $T_{term}$. During this intermittency, periods of SDC,
characterized by the stochastic creation and annihilation of spiral waves, are
 sandwiched between periods during which the system exhibits a small and 
stable number  of spiral waves. 
We demonstrate that the duration of these non-SDC periods can be prolonged by adjusting a model parameter, leading to a substantial increase in the average termination time.

\section{Methods}

\subsection{Electrophysiological model}
We use the Fenton-Karma (FK) model, a widely used cardiac model \cite{fenton1998vortex}, for our study.
The FK model consists of three variables: the variable $u$ that represents transmembrane potential, the fast gating variable $v$, and the slow gating variable $w$. 
These variables obey the following PDE and  ODEs:

\begin{align}
  \frac{\partial{u}}{\partial{t}} &  = D {\nabla^2}u - [J_{fi}(u,v) + J_{so}(u) + J_{si}(u,w)]  \label{eqn:pde}  \\ 
   \frac{\partial v}{\partial{t}} & = \frac{\Theta(u_c-u) (1-v)}{\tau_{v}^{-}(u)} - \frac{\Theta(u-u_c) \ v}{\tau^{+}_{v}}  \\ 
    \frac{\partial w}{\partial{t}} & = \frac{\Theta(u_c-u)(1-w)}{\tau_{-w}} - \frac{\Theta(u-u_c) w}{\tau_{+w}} \label{eqn:w} \\
  J_{fi}(u,v) & = -\frac{v}{\tau_d} \Theta(u-u_c) (1-u) (u-u_c)  \\
   J_{so}(u) & = -\frac{u}{\tau_o} \Theta(u_c-u) + \frac{1}{\tau_r} \Theta(u-u_c) \\
   J_{si}(u,w) & = -\frac{w}{2 \tau_{si}}(1 + \tanh[k(u-u_c^{si})])  \\
   \tau_{v}^{-}(u) & = \Theta (u_v-u) \ \tau_{-v2} + \Theta (u-u_v) \ \tau_{-v1}  
\end{align}
where $C_m$ ($\mu \rm{F}\, cm^{-2}$) is the membrane capacitance, the diffusive term  expresses the inter-cellular
coupling via gap junctions and diffusion constant $D$,  $J_{si}$ and $J_{so}$ are slow-inward and slow-outward currents, respectively,  $J_{fi}$ is the fast-inward  current, and   $\Theta$ represents the Heaviside function. 
The parameter values are listed in Table \ref{table:FK_Parameters}.   

\begin{table}
    \setlength{\tabcolsep}{10pt} 
    \renewcommand{\arraystretch}{1.05} 
    \begin{tabular}{||c|c||}
    \hline 
        {Parameter name} & {Parameter value}\\ [0.25ex] 
    \hline \hline
        $\tau_v^{+}$ & 3.3 ms \\
        $\tau_{-v1}$ & 12 ms \\ 
        $\tau_{-v2}$ & 2 ms \\   
        $\tau_{+w}$ & 1000 ms \\ 
        $\tau_{d}$ & 0.36 ms \\ 
        $\tau_{0}$ & 5 ms\\
        $\tau_{r}$ & 33 ms\\ 
        $\tau_{si}$ & 29 ms \\   
        $k$ & 15 \\ 
        $u_{c}^{si}$ & 0.7 \\ 
        $u_{c}$ & 0.13 \\    
        $u_{v}$ & 0.04 \\     
        \bm{$\tau_{-w}$} & 50-300 ms \\
    \hline
    \end{tabular}
    \caption{FK-model parameters used in our simulations. The parameter in bold face, $\tau_{-w}$, is varied within the indicated range.}
    \label{table:FK_Parameters}
\end{table}

\subsection{Simulation details}
We perform direct numerical simulations of the FK PDE model in a 2D homogeneous domain of size $N \times N$ with the no-flux boundary conditions. We employ a five-point stencil algorithm to solve the Laplacian and the forward Euler method for the time advancement.  We use $N=180$ with a spatial discretization of $\Delta x=0.025$ cm and a time step of $\Delta t=0.1$ ms, which, together with a diffusion coefficient $D = 0.0005 cm^2/ms$),  satisfies the von Neumann stability criterion.
\newline
We track spiral wave tips every $10$ms using a standard method 
that computes the phase at each instant of time and each grid point ~\cite{iyer2001experimentalist}.
By evaluating the line integral of the phase change around each grid point, we can identify a phase singularity (PS), which corresponds to the tip of a spiral wave.
In this study, we vary the time constant $\tau_{-w}$, which parameterizes the recovery time duration of $w$, the plateau current $J_{si}$, and hence the action potential duration (APD) of $u$.
It therefore determines the recovery time of an excitation, thereby controlling the ability of a tissue region to become excited again.
\newline
As our study is focused on SDC and intermittency, the different initial conditions for SDC were generated by starting with a dynamical state exhibiting multiple spiral waves and creating independent samples by adding uniformly distributed noise with an amplitude $10^{-5}$ to all three variables and at every grid point. These samples are then further evolved for a duration of $1 s$ to remove transient behavior, thereby finally creating multiple independent initial conditions. This procedure is repeated for each value of $\tau_{-w}$ considered
and simulations continue until all activity is terminated. 
Since the termination time varies for different values of $\tau_{-w}$ (see below), the number
of independent simulations ranges  
from 500 for $\tau_{-w}=50-185$ ms to 246 for $\tau_{-w}=200$ ms, 49 for $\tau_{-w}=275$, 7 for $\tau_{-w}=215,250$ ms and 8 for $\tau_{-w}=300$.

\begin{figure*}
	\includegraphics[scale=0.089]{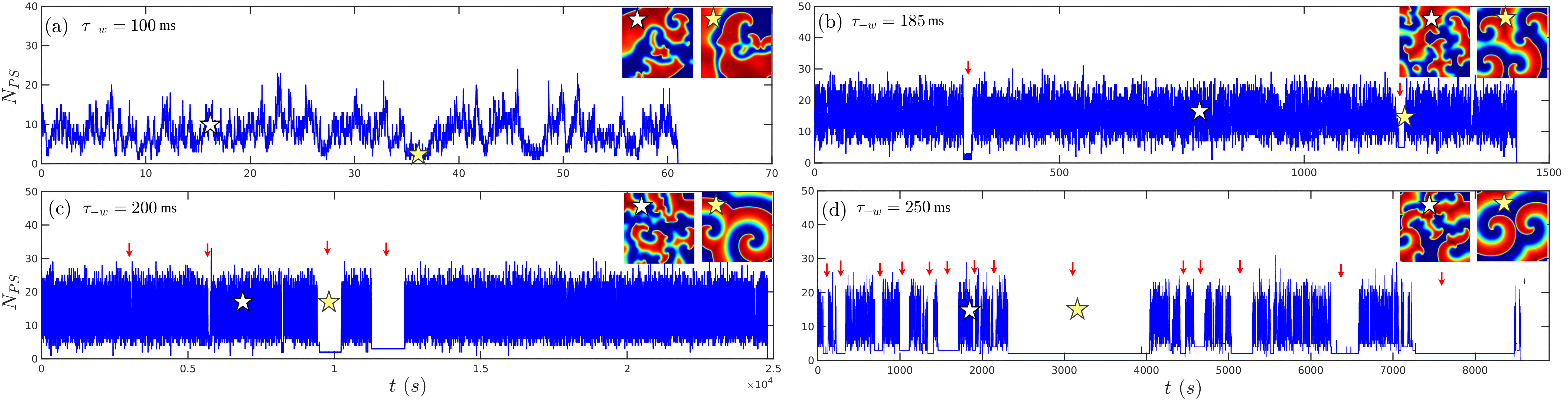}
	\caption{
		The number of phase singularities $N_{PS}$ as function of time for different values of $\tau_{-w}$.
		(a) Continuous SDC behavior  marked by stochastic fluctuations in $N_{PS}$ and an abrupt termination ($\tau_{-w} = 100$ ms). (b-d) SDC intervals, 
		intermittently interrupted by non-SDC intervals (indicated by 
		red arrows) during which $N_{PS}$
		remains stable. The length of these non-SDC intervals increases as $\tau_{-w}$
		increases.}
	\label{fig:fig1}
\end{figure*}

\section{Results}

\subsection{SDC with intermittency}
\label{sec:sec1}
For small values of $\tau_{-w}< 150$ms, the tissue recovers rapidly and a spiral wave becomes unstable close to its tip. As detailed in previous studies, this instability is due to a steep 
action potential duration restitution curve, which quantifies how the 
action potential duration depends on  the previous action potential duration 
and the time between excitations \cite{nolasco1968graphic,fenton2002multiple}.
As a result,  the model exhibits SDC and the number of 
spiral wave tips, $N_{PS}$,  fluctuates around an average number. 
Due to these fluctuations, the number of tips eventually reaches zero and SDC terminates.  This is shown in Fig.~\ref{fig:fig1}(a)), where we plot $N_{PS}$
as a function of time, rescaled by dominant period in the power spectrum, 
computed as $T_D=80$ ms.
This dominant period corresponds to the average period of a spiral wave in the 
SDC regime and is found to be largely independent of $\tau_{-w}$.
The insets show typical snapshots of the simulation, 
displaying the membrane potential using a color scale with red (blue) corresponding to high (low) values of $u$. These snapshots reveal 
unorganized dynamics and spiral waves that are continuously created and 
annihilated. 

For increasing values of $\tau_{-w} \geq 150$ ms, the instability of spiral waves no longer 
occur close to the tip as the restitution curve becomes flatter. 
Instead, the instability is characterized by so-called far-field breakup, which 
occurs far from the spiral wave tip \cite{fenton2002multiple}.
For these increasing values of $\tau_{-w}$ we find that the dynamics of the systems changes: 
instead of exhibiting continuous SDC,   the disordered state is intermittently interrupted
by a dynamical state characterized by a small and stable number of spiral waves 
(Fig.~\ref{fig:fig1}(b-d)). 
This non-SDC state is also demonstrated in the accompanying  
snapshots of $u$, which 
show a small number of spiral waves with a much larger spatial extent than 
the spiral waves during the SDC interval.
As the parameter is increased, the duration of these stable periods, indicated by 
red arrows, becomes 
longer and longer until only a single spiral wave persists. 
In our study, simulations with $\tau_{-w} = 300$  ms still exhibited periods of SDC, whereas all simulations with $\tau_{-w} \ge 330$  ms eventually resulted in a single stable spiral wave.
Thus, by varying a single parameter, the system's dynamics transition from completely disordered (SDC), to intermittently disordered, and finally to a fully ordered state.

\begin{figure}
\includegraphics[scale=0.125]{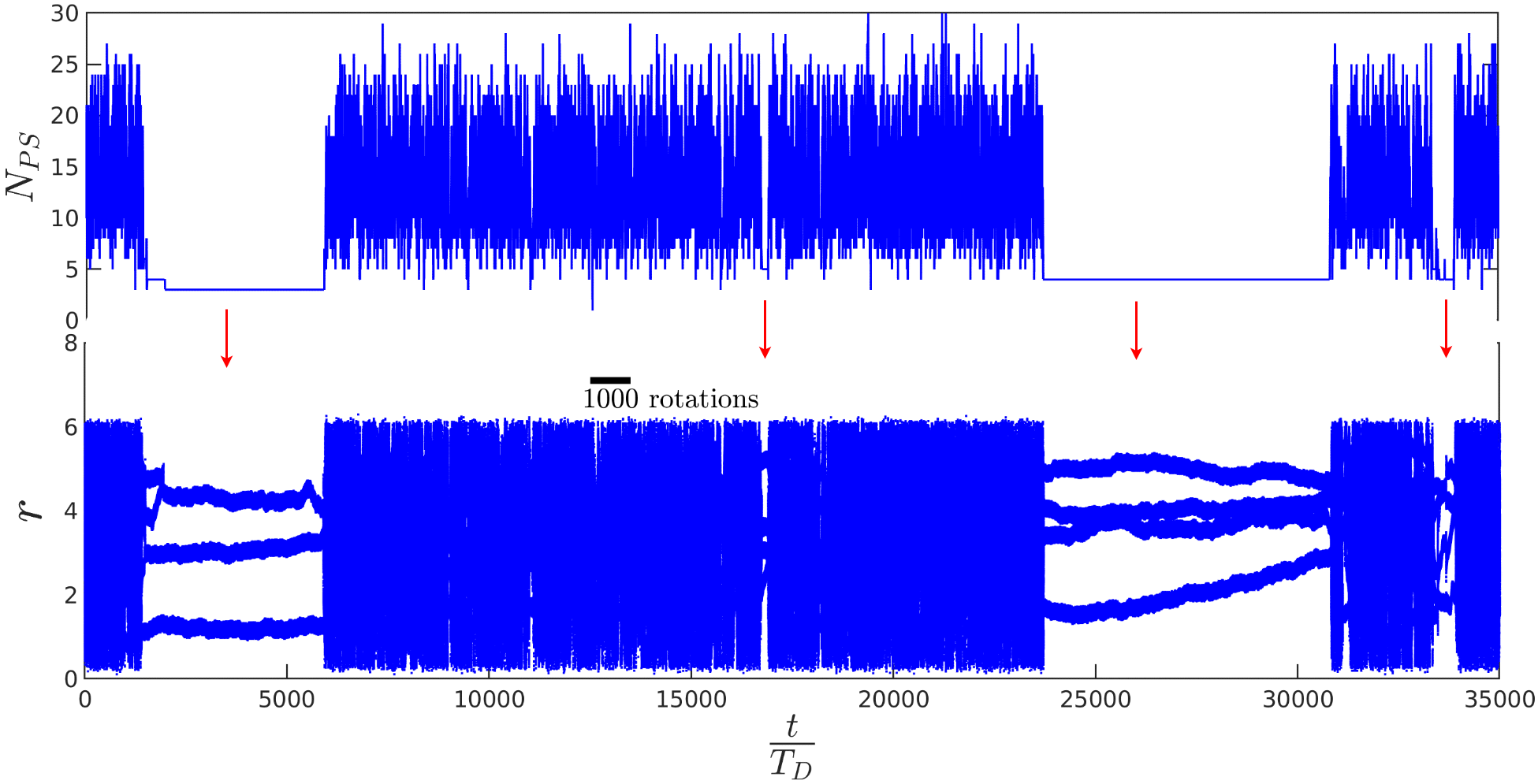}
\caption{Number of phase singularities (upper panel) and 
	distance of spiral wave tips (lower panel) as a function of time 
	($\tau_{-w}=200$ ms), highlighting the stability of spiral waves 
	during the non-SDC intervals (indicated by red arrows).
}
\label{fig:fig2}
\end{figure}

The intermittent dynamics is further exemplified in Fig.~\ref{fig:fig2}, which shows 
three non-SDC intervals, again indicated by red arrows, sandwiched between  SDC intervals. The lower panel shows, for the same time interval, the distance of the spiral wave tips to the lower left hand 
corner of the computational box, $r$. The first non-SDC interval contains 
a stable number of spiral waves (three) that are present for a duration 
of approximately 3000 $T_D$. In both the second, much shorter-lived non-SDC interval, and in the third non-SDC interval, lasting about 7000 $T_D$, 
four spiral waves are present.

\begin{figure}
\includegraphics[scale=0.224]{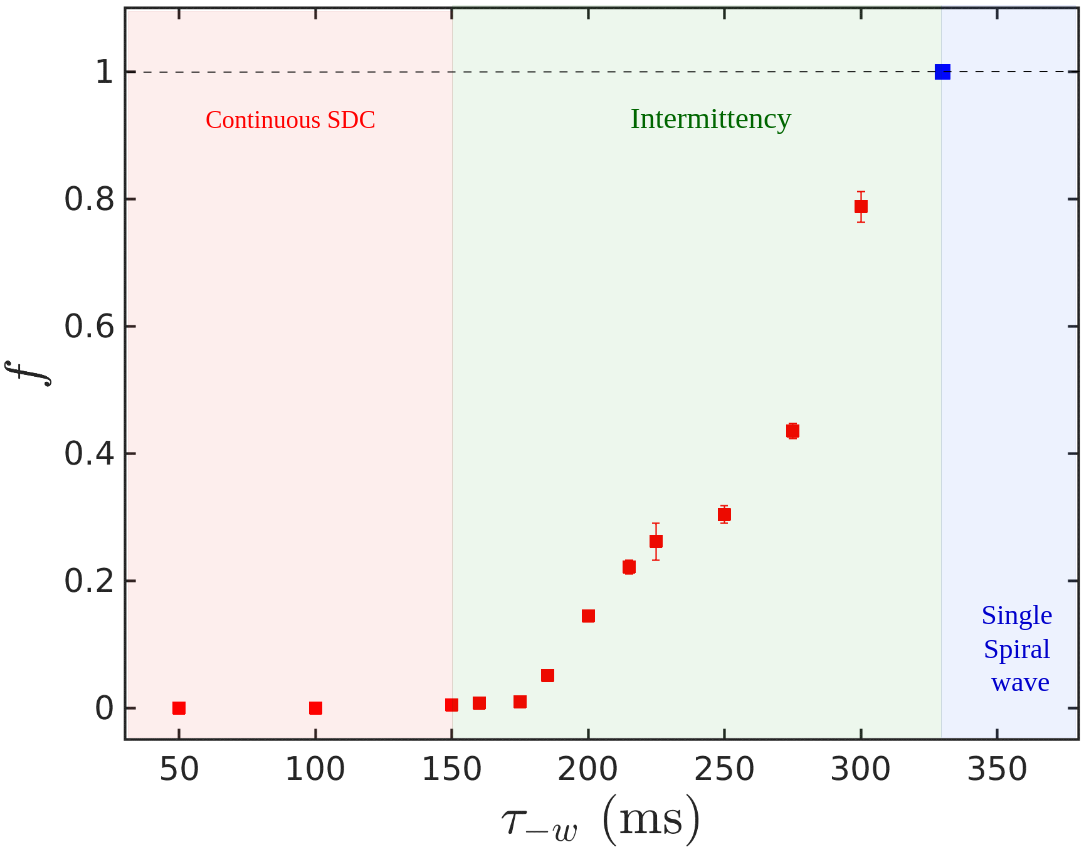}
\caption{Intermittency parameter $f$ as a function of $\tau_{-w}$ (red symbols) together with a blue symbol, representing the first parameter value for which all simulations 
resulted in a single spiral wave. The error bars are calculated using a bootstrap method with 10000 resamples. The different dynamical regimes of the model are color-coded.}
\label{fig:fig3}
\end{figure}

To quantify the extent of intermittency, we calculate the fraction of  time   the system spends in the  non-SDC state. 
For this, we define a non-SDC interval as any interval of length greater than 1.5 s, corresponding to 
at least 19 rotations,
in which the number of spiral wave tips remain stable. 
Next, we  identify all pairs of successive non-SDC  and SDC intervals and determine
their respective duration, $t_{non-SDC,i}$ and $t_{SDC,i}$. 
For each pair $i$ we calculate the  fraction of time spent in the non-SDC state as
$f_i=(t_{non-SDC,i})/(t_{non-SDC,i} + t_{SDC,i})$.
We then define the intermittency parameter, $f$, as the average of these individual fractions across all such pairs:
 \begin{eqnarray}
     f = \frac{1}{n} \sum_{i=1}^{n} f_i  
     \label{eqn:Intermittency}
 \end{eqnarray}
 where $n$ is the total number of non-SDC/SDC pairs.
Thus, $f=0$ corresponds to continuous SDC behavior with no intermittency and $f=1$ corresponds to a state with a spiral wave. 
In Fig.~\ref{fig:fig3} we plot $f$  as a function of $\tau_{-w}$ as symbols. Consistent with 
Fig.~\ref{fig:fig1}, we can 
identify three regions in this plot: a region corresponding to continuous SDC (pink region), 
a region corresponding to intermittency (green region), and a region corresponding 
to a single spiral wave (blue region). Clearly, within the 
intermittency region, the 
fraction of time spent in the non-SDC state gets larger as $\tau_{-w}$ increases.

\subsection{Distribution of spiral wave tips}
\label{sec:sec2}
\begin{figure*}
\includegraphics[scale=0.255]{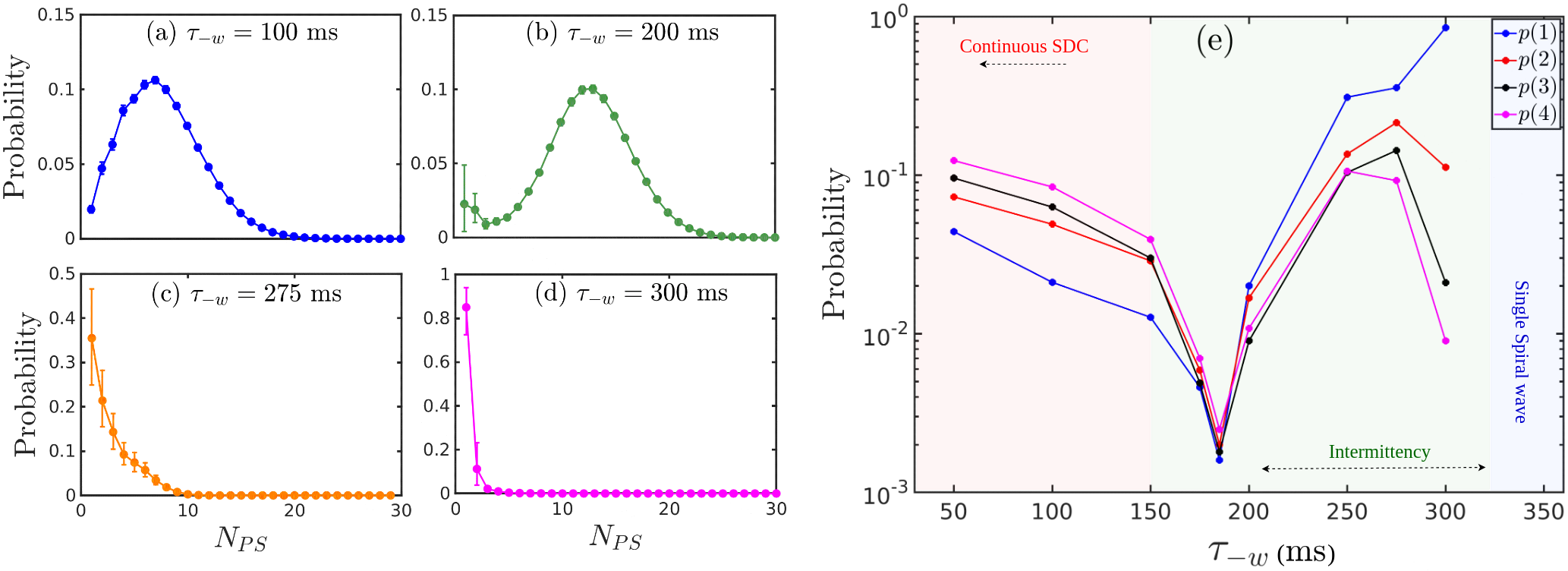}
\caption{(a-d): Probability distribution of $N_{PS}$  for different values of $\tau_{-w}$. Error bars represent  standard deviations  and solid lines correspond to mean values. 
	(e):  Probability of the system residing at $N_{PS}=1,2,3, \text{and} \ 4$ for different values of $\tau_{-w}$. }
\label{fig:fig4}
\end{figure*}

To further quantify the intermittency, we compute the probability distribution of 
spiral wave tips for different values of $\tau_{-w}$ (Fig.~\ref{fig:fig4}).
For values corresponding to continuous SDC (Fig.~\ref{fig:fig4}a), the distribution displays the 
characteristic bell-shaped curve discussed in previous work with the probability 
smoothly decreasing as the number of tips approaches one
\cite{vidmar2019extinction}. In this case, the mean termination time
is proportional to the inverse of the probability of having a single 
spiral tip in the domain ($p(N_{PS}=1$)) such that it decreases for increasing values of this probability \cite{vidmar2019extinction}.  
For increasing values of $\tau_{-w}$ in the intermittency region, however, the 
distribution starts to exhibit a `bump' for small values of $N_{PS}$, reflecting the 
non-SDC intervals with a several quasi-stable (since they eventually de-stabilize) spiral waves (Fig.~\ref{fig:fig4}b).
For even larger values of   $\tau_{-w}$ the distribution becomes dominated
by small values of $N_{PS}$ (Fig.~\ref{fig:fig4}c \& d), indicating that the dynamics
is mainly governed by non-SDC intervals. Furthermore, since 
for large values of  $\tau_{-w}$ a single spiral wave is stable, 
the 
distribution for $\tau_{-w}=300$ ms becomes highly peaked at  $N_{PS}=1$
(see Fig.~\ref{fig:fig4}d).

The increase in the probability of having a small number of tips is also 
shown in Fig.~\ref{fig:fig4}(e) where we plot $p(N_{PS})$ for
$N_{PS}=1-4$ as a function of $\tau_{-w}$.
These probabilities decrease for values of $\tau_{-w}$ up to 
$\tau_{-w}=185$ ms. This decrease becomes more pronounced when 
the intermittency regime is entered. 
For $\tau_{-w}>185$ ms, the probabilities increase, reflecting the 
dominance of non-SDC intervals. 
Note that for large values of $\tau_{-w}$ and before entering the 
regime of a single spiral wave, $p(N_{PS}=1)$  continues to increase
at the expense of $p(N_{PS} \ne 1)$.

\subsection{SDC with intermittency increases the mean termination time}
\begin{figure*}
\includegraphics[scale=0.265]{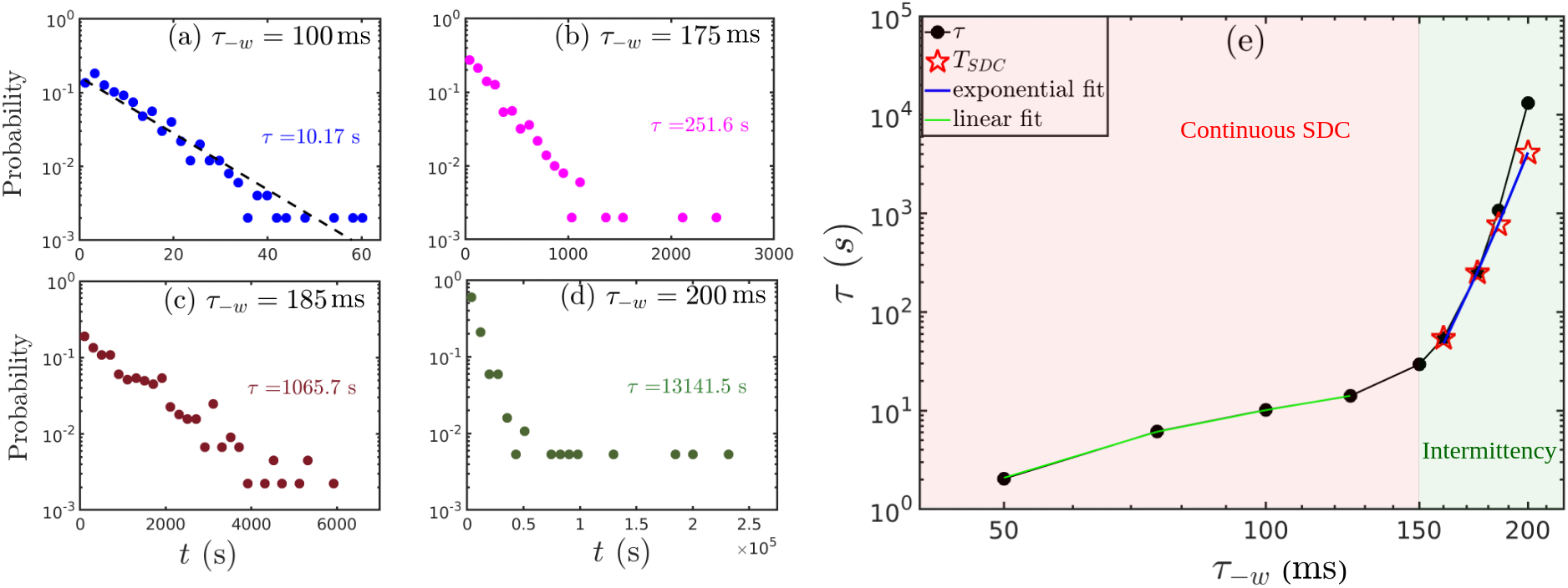}
\caption{ (a)-(d) Semi-logarithmic plot of the 
	probability distribution of termination times for increasing values of $\tau_{-w}$.
    The dashed line in (a) represents an exponential fit. (e) $T_{term}$ as a function of  $\tau_{-w}$ (solid symbols) and 
	the time spent in the SDC state (open symbols).
In the continuous SDC parameter regime, the increase in $\tau$ is linear, as shown by the linear fit (green line). 
In the intermittency parameter regime the time spent in the SDC state ($\tau_{SDC}$, red stars) increases exponentially as 
demonstrated by the exponential fit (blue line). 
Notice the growing difference between  $T_{SDC}$ (red stars) and $T_{term}$ (black filled circles) for
increasing values of $\tau_{-w}$, implying that non-SDC regions exert substantial influence on the 
mean termination time.}
\label{fig:fig5}
\end{figure*}

The  appearance of intermittent  non-SDC intervals implies that the 
mean termination time can no longer be  approximated using 
$p(N_{PS}=1)$. Instead, these non-SDC intervals, containing a few spiral wave tips, will increase the mean termination time. 
To quantify this increase and the dependence of $T_{term}$ on   $\tau_{-w}$, we compute the 
termination time of a large number of simulations ($>$100) that are started
with independent initial conditions. 
Examples of the resulting distribution of termination times are presented 
in Fig.~\ref{fig:fig5} for a value of $\tau_{-w}$ in the continuous SDC
regime (a) and for values in the intermittent regime (b-d). In the continuous SDC regime, 
the distribution is exponential, consistent with previous work \cite{vidmar2019extinction,dharmaprani2019renewal}, 
as can also be seen by the exponential fit (dashed line in Fig.~\ref{fig:fig5}a). 
In the intermittent
regime, however, the termination times exhibit a long tail, corresponding to 
simulations with multiple or long-lasting non-SDC intervals. 

Using these termination times, we compute the mean termination time 
as a function of $\tau_{-w}$ (solid circles Fig.~\ref{fig:fig5}e). 
In the continuous SDC regime, we find that $T_{term}$  increases linearly with $\tau_{-w}$. 
Once $\tau_{-w}$ enters the intermittent regime, however, 
$T_{term}$   becomes strongly non-linear and 
sharply increases by almost 
three orders of magnitude between $\tau_{-w}=150$ ms and 
$\tau_{-w}=200$ ms. 
 Note that
due to the computational expense, we are not able 
to compute $T_{term}$ for larger values of $\tau_{-w}$. 
Also note that  for  $\tau_{-w} \ge 330$ ms only a single spiral wave exists,  implying $\tau \rightarrow \infty$.

For values of $\tau_{-w}$ in the intermittency regime, we  also compute the average amount of time spent in the SDC state before termination, $T_{SDC}$ (open red stars in Fig.~\ref{fig:fig5}e).
For values up to $\tau_{-w}=175$ ms, this time is almost indistinguishable
from $T_{term}$, indicating that the non-SDC intervals are not significantly
affecting the mean termination time. 
Of course, this is consistent with the small value of $f$ for these choices of
$\tau_{-w}$ (Fig.~\ref{fig:fig3}). In other words, 
the sharp increase in $T_{term}$ is not due to the system spending 
more and more time in the non-SDC state. Instead, it is due to the significant reduction in the probability of having a small number of tips, as shown in Fig. ~\ref{fig:fig4}e.

\begin{figure*}
\includegraphics[scale=0.205]{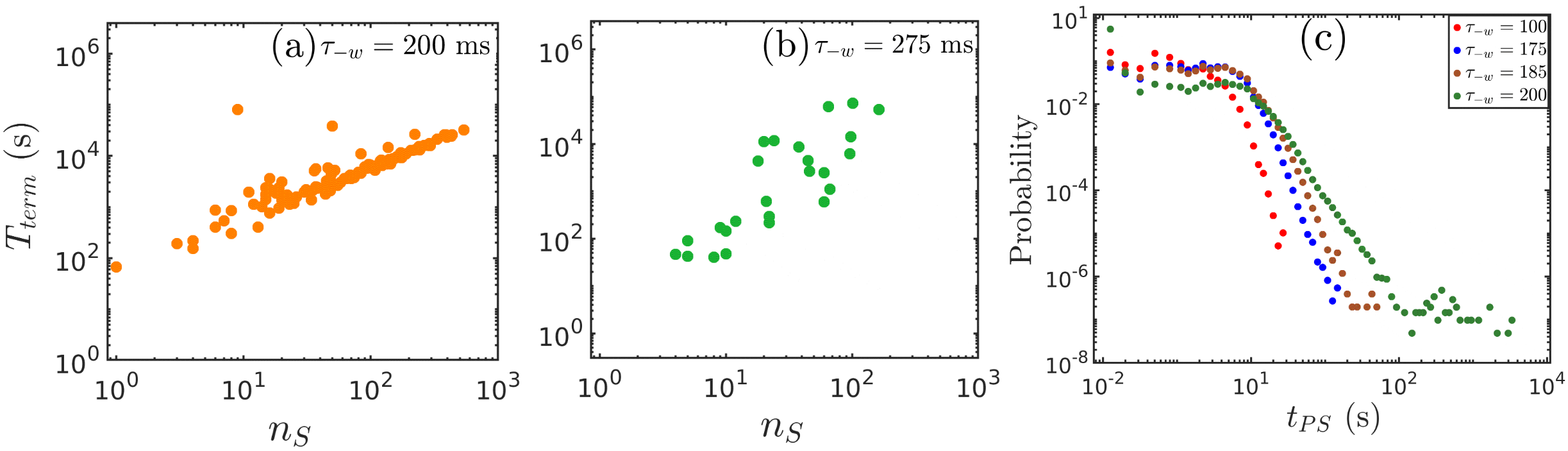}
\caption{(a-b) Mean termination time as a function of the number of non-SDC intervals for
$\tau_{-w}=200$ ms (a) and $\tau_{-w}=275$ ms (b). Each symbol represent a simulation in which 
SDC terminates. 
(c) Probability distribution of spiral wave lifetimes for different values of $\tau_{-w}$. 
    As $\tau_{-w}$ increases, longer lives spiral waves become more common.
}
\label{fig:fig6}
\end{figure*}

The correlation between the appearance of intermittent non-SDC intervals and the termination time $T_{term}$ can be further illustrated by examining multiple terminating simulations.
For this, we  compute the
 number of intermittent non-SDC intervals, $n_S$, for each separate terminating simulation. a
As shown in Fig.~\ref{fig:fig6}a\&b for two different values of $\tau_{-w}$, $n_S$ exhibits a clear positive correlation with $T_{term}$. Thus, 
for a fixed value of  $\tau_{-w}$, simulations with more intermittent intervals tend to terminate later. 
\newline

\begin{figure}
	\includegraphics[scale=0.232]{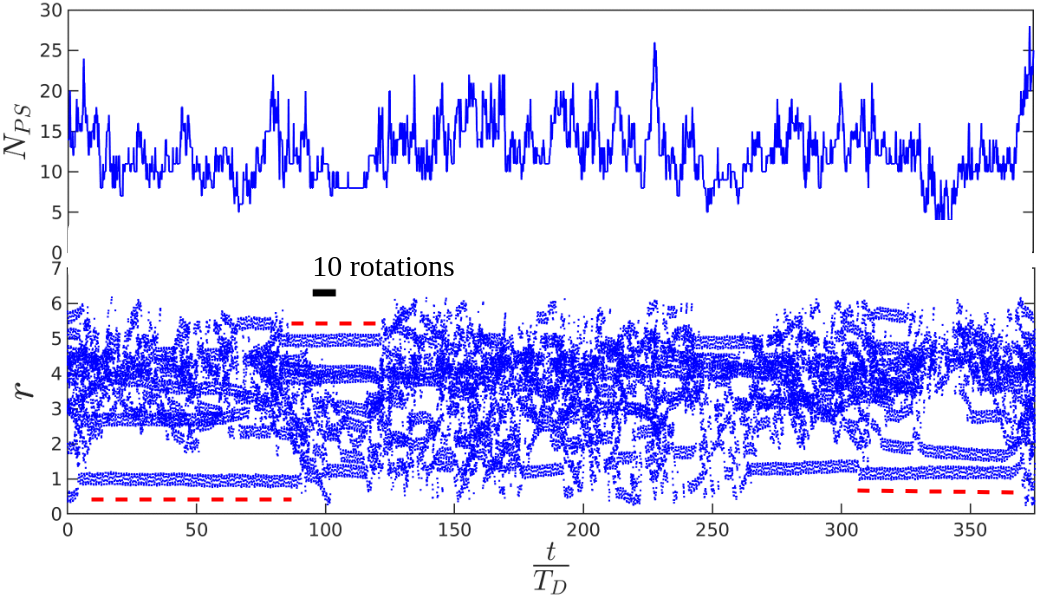}
	\caption{Number of spiral wave tips (upper panel) and 
		distance of tips to corner of computational domain (lower panel) as a function of normalized time ($\tau_{-w}=200$ ms). Quasi-stable spiral waves that co-exist with SDC are 
		indicated by dashed red lines.  }
	\label{fig:fig7}
\end{figure}

To further verify that the presence of quasi-stable spirals increases $T_{term}$, we calculate the lifetime of each spiral wave tip, $t_{PS}$. 
The probability distribution of these liftetimes is presented in  Fig.~\ref{fig:fig6}c
for different values of  $\tau_{-w}$. 
For small values of $\tau_{-w}$ that correspond to continuous SDC (red symbols) , the lifetimes
are at most several seconds. 
As $\tau_{-w}$ increases, however, the probability distribution shifts to larger and larger values of 
$t_{PS}$ due to the appearance of long-lasting, quasi-stable spiral waves. For example,  
for $\tau_{-w}=200$ ms spiral waves can last for more than $10^3$ s, 
leading to large termination times.
\newline
Finally, we want to point out that it is also possible to have intermittently 
present quasi-stable spiral waves in part of the computational domain while the 
remainder of the domain exhibit SDC. 
This is illustrated in  Fig. \ref{fig:fig7} where we plot $N_{PS}$ (upper panel)
and $r$ as a function of normalized time. While $N_{PS}$ fluctuates during this time
interval, $r$ for some spiral waves clearly remains constant, as indicated by the 
dashed red lines. In other words, quasi-stable spiral waves co-exist with 
stochastically formed spiral waves and these waves were not considered when 
computing $f$ (Fig. \ref{fig:fig3}) nor when determining the data presented in Fig. \ref{fig:fig6} (a) \& (b). 
Examining our simulation results reveals that 
this co-existence becomes more prominent for $\tau_{-w}\geq 200$ ms.

\section{Concluding Remarks}
In this paper, we show that cardiac models can exhibit intermittent dynamics during which 
periods of SDC are alternated with periods of a more ordered dynamical state, characterized 
by a small number of quasi-stable spiral waves. 
This intermittency manifests itself when we vary a single parameter, $\tau_{-w}$ in the FK model, 
representing the  recovery time duration of one of the model variables. 
In fact, by varying this parameter across a range of values, we show that the system transitions from SDC to intermittent dynamics, and ultimately to a single spiral wave.
The dynamics can be quantified using an intermittency parameter $f$, which measures the 
fraction of time the system was in the more ordered, non-SDC state. 
This parameter shows a  transition from 0, representing continuous SDC, to 1, corresponding 
to a single spiral wave, over a limited range of $\tau_{-w}$.
As a result of the appearance of the intermittent, non-SDC intervals, the mean termination time 
of  wave activity increases dramatically, especially for values of $\tau_{-w}$ that 
approach the regime of a single spiral wave. 
Thus, similar to changing the area of the computational domain,  altering this parameter can 
significantly lengthen the mean termination time. 
We should comment here that the intermittent behavior is not restricted to the particular parameter change, parameter set or  model under consideration. We have observed similar behavior when  changing
another parameter ($\tau_{+w}$), using other parameter sets of the FK model, and in a generic two-variable model ~\cite{aliev1996simple}.

The fact that the termination time sharply increases as $\tau_{-w}$ increases and enters the intermittency 
regime is perhaps not unexpected. After all, for small values of $\tau_{-w}$, the 
system exhibits SDC, which has a finite lifetime (see Fig. \ref{fig:fig5}e). On the other hand, 
for large values of $\tau_{-w}$, the system becomes ordered and exhibits a single spiral wave with 
a termination time that becomes infinite. In the range of parameter values between these two 
extremes, the termination time increases rapidly due to the appearance of longer and longer
non-SDC intervals (see Fig. \ref{fig:fig6}).

Intermittency is a fundamental phenomenon in dynamical systems where the system alternates between periods of regular behavior and bursts of chaotic or disordered activity.
It was first described in dissipative systems, where it was shown to be a 
key route to chaos \cite{pomeau1980intermittent}. Intermittent dynamics can be 
found in biological and biochemical systems \cite{cabrera2002off, harnos2000scaling, de1996intermittency}
and is wide-spread in physical systems. It has been particularly well studied in fluid dynamics, where varying the Reynolds number can cause a transition from laminar flow to intermittent and ultimately fully turbulent behavior \cite{moxey2010distinct,barkley2016theoretical}.
Thus, our control parameter, $\tau_{-w}$, plays a role analogous to the Reynolds number in fluid flow, with laminar flow corresponding to a single spiral wave and turbulent behavior corresponding to SDC.
However, contrary to fluid flow,  SDC can terminate, resulting in the complete cessation of 
all activity. 

The observed intermittency may have  clinically relevance for Afib. 
In many patients, Afib is initially intermittent with periods of fibrillation 
alternating with periods of normal sinus rhythm. While our simulations reveal an alternation between SDC, thought to underlie AFib, and more organized non-SDC states, it is possible that, in the presence of periodic pacing as occurs in the real heart, the non-SDC state becomes even more organized and ultimately transitions into sinus rhythm. Investigating such a scenario would require simulating 
SDC and its intermittency using realistic cardiac geometries that include periodic pacing from the 
sino-atrial node. In future work, we plan to address this problem.  In addition to this extension, we
also plan to extend our intermittency study to other cardiac models, including more detailed ones. 
This will allow us to determine whether specific ion channels contribute to the emergence of intermittency, possibly identifying therapeutic targets that can shorten the duration of 
Afib in patients.

\newpage

\section*{Acknowledgments}
We thank Akhilesh Kumar Verma and Sebastian Echeverria-Alar for helpful discussions.
This work was supported by National Institutes of Health Grant R01 HL122384.
\newline

\end{document}